\documentclass[reprint,superscriptaddress,nofootinbib,twocolumn,amsmath,amssymb,aps,pra,longbibliography]{revtex4-1}

\usepackage{graphicx}
\usepackage{dcolumn}
\usepackage{bm}
\usepackage[colorlinks]{hyperref}
\usepackage{multirow}

\begin{document}

\preprint{APS/123-QED}

\title{Controllable transitions among phase-matching conditions in a single nonlinear crystal}

\author{Zi-Qi Zeng}
\affiliation{Hubei Key Laboratory of Optical Information and Pattern Recognition, Wuhan Institute of Technology, Wuhan 430205, China}

\author{Shi-Xin You}
\affiliation{Hubei Key Laboratory of Optical Information and Pattern Recognition, Wuhan Institute of Technology, Wuhan 430205, China}

\author{Zi-Xiang Yang}
\affiliation{Hubei Key Laboratory of Optical Information and Pattern Recognition, Wuhan Institute of Technology, Wuhan 430205, China}

\author{Chenzhi Yuan}
\affiliation{Hubei Key Laboratory of Optical Information and Pattern Recognition, Wuhan Institute of Technology, Wuhan 430205, China}

\author{Chenglong You}
\affiliation{Quantum Photonics Laboratory, Department of Physics \& Astronomy, Louisiana State University, Baton Rouge, LA 70803, USA}

\author{Rui-Bo Jin}
\email{jin@wit.edu.cn}
\affiliation{Hubei Key Laboratory of Optical Information and Pattern Recognition, Wuhan Institute of Technology, Wuhan 430205, China}

\date{\today}

\begin{abstract}
Entangled photon pairs are crucial resources for quantum information processing protocols. Via the process of spontaneous parametric down-conversion (SPDC), we can generate these photon pairs using bulk nonlinear crystals. Traditionally, the crystal is designed to satisfy specific type of phase-matching condition. Here, we report controllable transitions among different types of phase-matching in a single periodically poled potassium titanyl phosphate (PPKTP) crystal. By carefully selecting pump conditions, we can satisfy different phase-matching conditions. This allows us to observe first-order type-II, fifth-order type-I, third-order type-0, and fifth-order type-II SPDCs. The temperature-dependent spectra of our source were also analyzed in detail. Finally, we discussed the possibility of observing more than nine SPDCs in this crystal. Our work not only deepens the understanding of the physics behind phase-matching conditions, but also offers the potential for a highly versatile entangled biphoton source for quantum information research.
\end{abstract}

\maketitle

\section{Introduction}
Quantum light sources, including single-photon sources and entangled photon sources, are fundamental resources for the study of quantum information processing \cite{Anwar2021,Zhang2022PI}.One of the most widely used methods for preparing quantum light sources is spontaneous parametric down-conversion (SPDC) in a nonlinear optical crystal \cite{Christ2013, Zhang2021a}.In an SPDC process, a pump photon interacts with a nonlinear crystal and is converted into a biphoton, which is a pair of correlated photons usually referred to as the signal and idler. The biphotons can be further engineered to prepare heralded single photon source or entangled photon source. The SPDC process can be engineered in different degrees of freedom, such as space, time, frequency, polarization, and phase \cite{Morrison2022, Zhu2023}.

According to the polarization of the pump, signal and idler photons, the phase-matching conditions in nonlinear SPDC can be classified into three types:
Type-0:  H$\to$HH, or V$\to$VV; Type-I:  H$\to$VV, or V$\to$HH;Type-II: H$\to$HV, or V$\to$HV.Here, ``H'' and ``V'' represent the horizontal and vertical polarizations respectively \cite{Dmitriev2013}. In the type-0 and type-I cases, the signal and idler photons have the same polarization, resulting in broad spectral distributions and narrow temporal distributions. This feature of broad spectra is advantageous in quantum metrology \cite{Nielsen2023, Reisner2022}, quantum optical coherence tomography \cite{Hayama2022}, and quantum spectroscopy \cite{Tabakaev2021, Chen2021}. Furthermore, the type-0 matching condition has the highest effective nonlinear coefficient among all three nonlinear phase-matching conditions \cite{Dmitriev2013}. In the type-II cases, the signal and idler photons usually have much narrower spectra and have been widely used to perform quantum communication, computation, and measurement \cite{Yin2017, Zhong2020, Lyons2018,Guo2023OE}. Another important feature of the type-II matching condition is that the joint spectral distribution of biphotons can be engineered according to the group velocity differences \cite{Edamatsu2011}. As a result, one can prepare photon pair sources with engineerable frequency correlation \cite{Graffitti2020PRL, Morrison2022, Zhu2023}.

Traditionally, one crystal is usually designed to satisfy only one specific phase-matching condition for quantum information processing protocols. For different protocols, one may need different crystals supporting different phase-matching conditions. Previous works have demonstrated the possibility to realize two SPDC processes in a single PPKTP crystal. These studies were limited to only type-II and type-0 phase-matching conditions, and the detailed properties of such sources are unclear \cite{Lee2012, Steinlechner2014, Chen2009, Laudenbach2017}. In this work, we demonstrate controllable transitions among all types of phase-matching in a single nonlinear crystal. Specifically, we find that a periodically poled potassium titanyl phosphate (PPKTP) with a poling period of 10 $\mu$m can realize all three types of phase-matching conditions simultaneously at pump wavelengths around 405 nm. We observed first-order type-II (at 404.3 nm), fifth-order type-I (at 404.3 nm), third-order type-0 (at 408.8 nm), and fifth-order type-II (at 315 nm) phase-matching conditions. More importantly, the design principles, detailed spectra and tuning curves in our sources are revealed for future references. Our work may provide a versatile biphoton source for diverse tasks in quantum information technology.

\begin{figure*}[!htbp]
\centering
\includegraphics[width=0.85\textwidth]{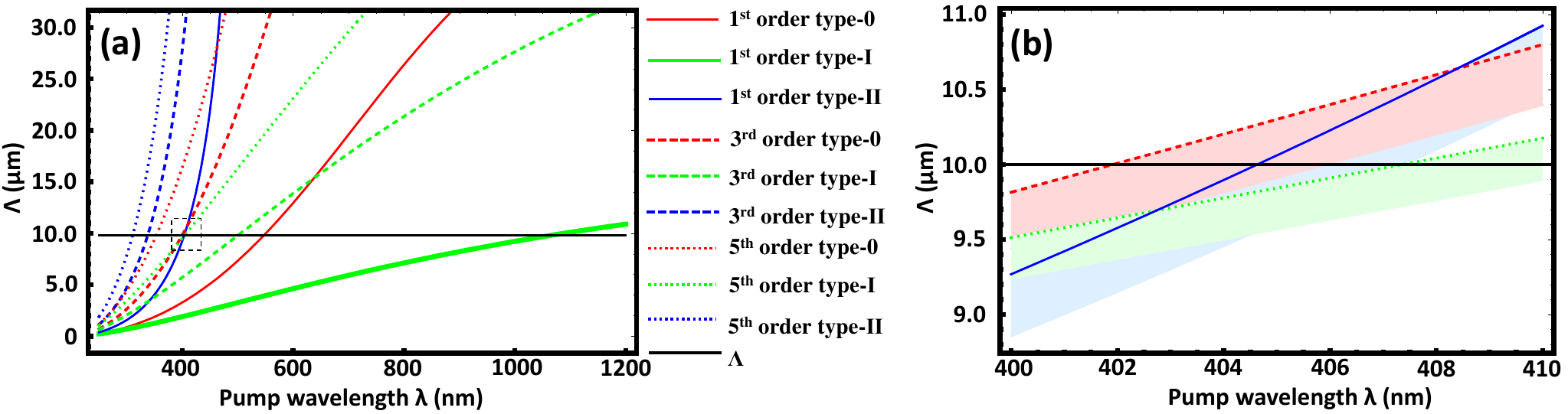}
\caption{The calculated poling period of the PPKTP crystal as a function of pump wavelength under different phase-matching conditions. The black horizontal line represents the 10 $\mu$m poling period. In this calculation, we assume the degenerate case of $\lambda_s=\lambda_i=2\lambda_p$, and the temperature of the crystal is set at 25 $^{\circ} \mathrm{C}$. As shown in (a), three phase-matching conditions are satisfied near $\lambda_p=405$ nm. In (b), we show a zoomed-in plot for pump wavelengths between 400 nm and 410 nm. Here, these shadowed regions represent the temperature tuning range of each phase-matching conditions. For each region, the upper bound and lower bound is at 25 $^{\circ} \mathrm{C}$ and 150 $^{\circ} \mathrm{C}$, respectively}
\label{Fig1}
\end{figure*}

\begin{table*}[!ht]
\centering
\begin{tabular}{c|c|c|c|c|c|c|c} 
\hline\hline
Poling order                       & Phase-matching & Polarization                         & Pump $(\mathrm{nm})$ & Signal $(\mathrm{nm})$ & Idler $(\mathrm{nm})$ & $d(z)(\mathrm{pm} / \mathrm{V})$                    & $\Lambda(\mu \mathrm{m})$  \\ 
\hline
\multirow{3}{*}{$1^{\text {st }}$} & Type-0         & $\mathrm{Z} \rightarrow \mathrm{ZZ}$ & 550.82               & 1101.65                & 1101.65               & $2 \mathrm{~d}_{33} / \pi=10.76$                      &                            \\ 
\cline{2-7}
                                   & Type-I         & $\mathrm{Z} \rightarrow \mathrm{YY}$ & 1070.91              & 2141.82                & 2141.82               & $2 \mathrm{~d}_{32} / \pi=2.77$                       &                            \\ 
\cline{2-7}
                                   & Type-II        & $\mathrm{Y} \rightarrow \mathrm{YZ}$ & 404.63               & 809.27                 & 809.27                & $2 \mathrm{~d}_{24} / \pi=2.32$                       &                            \\ 
\cline{1-7}
\multirow{3}{*}{$3^{\text {rd }}$} & Type-0         & $\mathrm{Z} \rightarrow \mathrm{ZZ}$ & 401.92               & 803.84                 & 803.84                & $2 \mathrm{~d}_{33} / 3 \pi=3.59$                     & \multirow{3}{*}{10}        \\ 
\cline{2-7}
                                   & Type-I         & $\mathrm{Z} \rightarrow \mathrm{YY}$ & 505.21               & 1010.41                & 1010.41               & $2 \mathrm{~d}_{32} / 3 \pi=0.92$                     &                            \\ 
\cline{2-7}
                                   & Type-II        & $\mathrm{Y} \rightarrow \mathrm{YZ}$ & 338.73               & 677.45                 & 677.45                & $2 \mathrm{~d}_{24} / 3 \pi=0.77$                     &                            \\ 
\cline{1-7}
\multirow{3}{*}{$5^{\text {th }}$} & Type-0         & $\mathrm{Z} \rightarrow \mathrm{ZZ}$ & 354.97               & 709.94                 & 709.94                & $2 \mathrm{~d}_{33} / 5 \pi=2.15$                     &                            \\ 
\cline{2-7}
                                   & Type-I         & $\mathrm{Z} \rightarrow \mathrm{YY}$ & 407.37               & 814.73                 & 814.73                & $2 \mathrm{~d}_{32} / 5 \pi=0.55$                     &                            \\ 
\cline{2-7}
                                   & Type-II        & $\mathrm{Y} \rightarrow \mathrm{YZ}$ & 311.43               & 622.86                 & 622.86                & $2 \mathrm{~d}_{24} / 5 \pi=0.46$                    &                            \\
\hline\hline
\end{tabular}
\end{table*}

\section{Theory}
In the process of an SPDC, the energy conservation law 
is satisfied and is given as
\begin{equation}\label{eq1}
\omega_p=\omega_s+\omega_i,
\end{equation}
where $\omega$ is the angular frequency and the subscripts $p$,  $s$, and  $i$ denote the pump, signal, and idler, respectively. The momentum conservation law is also satisfied and can be expressed in the form of phase-matching function (PMF). For quasi-phase-matched (QPM) crystals, the PMF is given by 
\begin{equation}\label{eq2}
k_p-k_s-k_i \pm m\frac{2\pi}{\Lambda}=0,
\end{equation}
where $k_j(j=p, s, i)$ is the wave vector, $\Lambda$ is the poling period, and $m=1,3,5...$ is the poling order.
\begin{figure*}[!ht]
\centering
\includegraphics[width=0.85\textwidth]{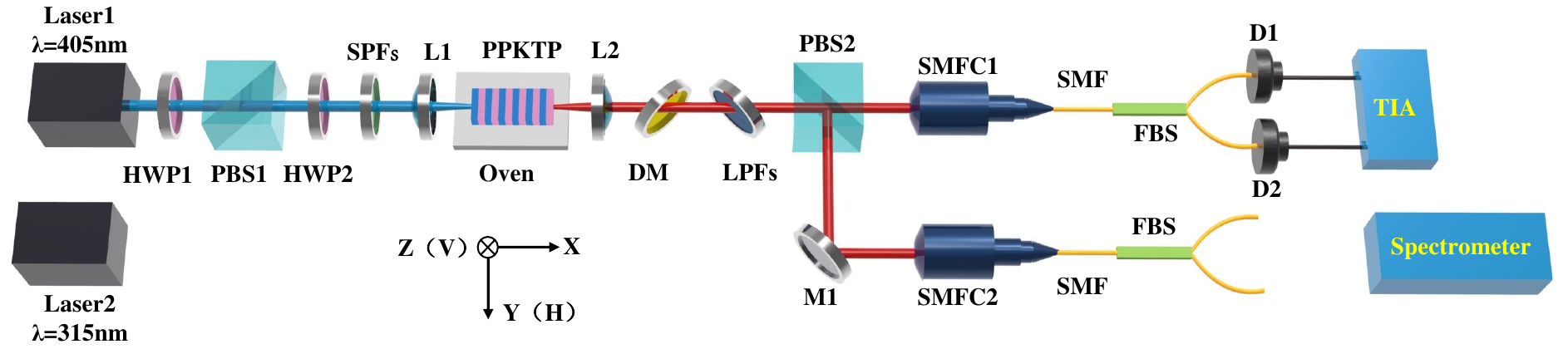}
\caption{The experimental setup for observing controllable transitions among different SPDC processes. Here, we used two lasers at 405 nm and 315 nm. We control the power and polarization of the pump beam, and we also control the temperature of the PPKTP crystal. The properties of these SPDC processes are characterized using a time interval analyzer (TIA) and a single-photon level spectrometer. HWP: half-wave plate, PBS: polarizing beam splitter, SPFs: short-pass filters, L: lens,  DM: dichroic mirror, LPFs: long-pass filters, M: Mirror, SMFC: single-mode-fiber coupler, SMF: single-mode fiber, FBS: fiber beamsplitter, D: detector}
\label{Fig2}
\end{figure*}

In a PPKTP crystal, the photons are usually designed to propagate along the x-direction but polarize in the y- and z-directions. Here, we assign y-direction to H polarization and z-direction to V polarization.
Using the Sellmeier equations of KTP from Ref.\cite{Fan1987} for refractive index $n_y$, Ref.\cite{Fradkin1999} for $n_z$, and Ref.\cite{Emanueli2003} for temperature-dependent dispersion, the PMF in Eq.\,\ref{eq2} can be calculated.

In Fig.\,\ref{Fig1}(a), we show the relation between pump wavelength and poling period with nine different combinations of poling order and phase-matching type. 
{Tab.\,\ref{Tab:1} shows the specific wavelength values of the cross points in  Fig.\,\ref{Fig1}(a).}
Notably, there are three SPDC processes near the wavelength of 405 nm. We inspect these three phase-matching conditions further in Fig. \ref{Fig1}(b) by zooming in the wavelength range between 400 nm and 410 nm. We can confirm that the cross points are at the pump wavelengths of 401.9 nm ($3^\textrm{rd}$-order type-0), 404.6 nm ($1^\textrm{st}$-order type-II), and 407.4 nm ($5^\textrm{th}$-order type-I). This feature allows us to achieve a controllable transition among type-0, type-I, and type-II phase matching condition by tuning the pump wavelength in a small range between 401 nm and 408 nm. Moreover, we can adjust the cross points by tuning the temperature of the PPKTP crystal. Finally, we note that our calculation indicates the existence of another experimentally implementable SPDC process of $5^\textrm{th}$-order type-II at 311.4 nm, as shown in Fig.\,\ref{Fig1} (a).

The effective nonlinear coefficient is an important parameter for QPM crystals, which could be different for phase-matching conditions of different types and orders. We define $d_{\text{eff}}$ as the effective nonlinear coefficient of KTP and $d(z)$ as the nonlinear coefficients of PPKTP at the position of $z$. $d(z)$ is a square-wave function \cite{boyd2020nonlinear, Niu2023, niu2021}:
\begin{equation}\label{eq3}
d(z)=d_{\text{eff}} \mathrm{sign}[\cos (2 \pi z / \Lambda)].
\end{equation}
$d(z)$ can also be expressed by Fourier series:
\begin{equation}\label{eq4}
d(z)=d_{\text{eff}}\sum_{m=-\infty}^{\infty} G_m \exp (i{k_m}z),
\end{equation}
 where $k_{m}=2 \pi m / \Lambda$, and  $G_{m}$ can be expressed as 
 \begin{equation}\label{eq5}
G_{m}=(\frac{2}{m\pi})\sin(\frac{m\pi}{2}). 
 \end{equation}
Intuitively speaking, $k_{m}$ is the spatial frequency of the grating structure of the PPKTP crystal, and $G_{m}$ is the coefficient of its eigenmode $\exp (i{k_m}z)$. The detailed derivation of $d(z)$ and $G_{m}$ can be found in the Appendix.

\section{Experiment and results}

%
%
\begin{figure*}[!ht]
\centering
\includegraphics[width= 0.95\textwidth]{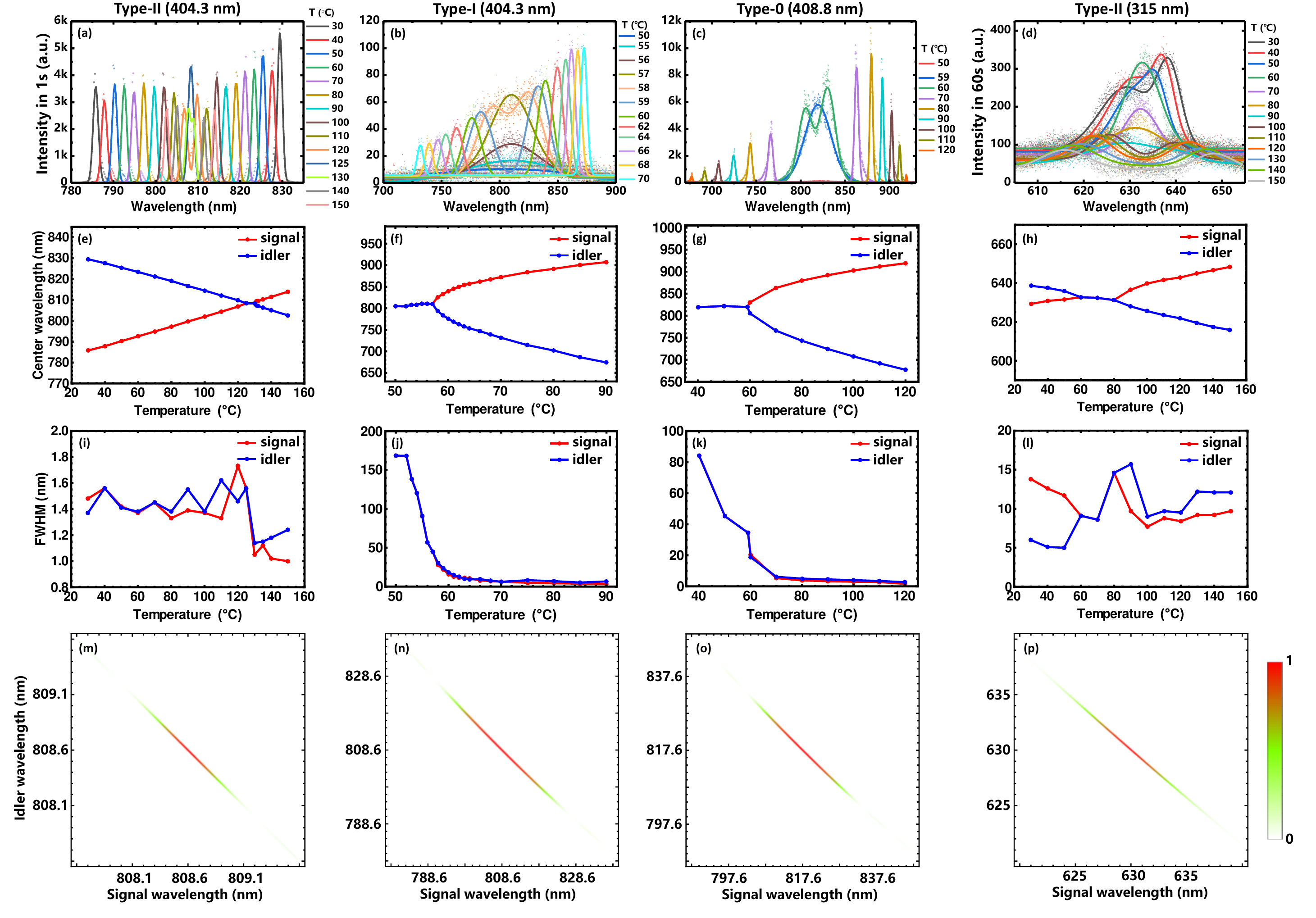}
\caption{The measured properties of our versatile biphoton source. (a-d): The spectra of the biphotons for different crystal temperatures. (e-h) The tuning curve of the four observed SPDCs. Here, we report the center wavelengths of the signal and idler photons at different temperatures. (i-l) The FWHM of the signal and idler photons at different temperatures. We note that the values in (e-l) were obtained by fitting each spectres to a Gaussian distribution.{(m-p) The simulated joint spectral intensities (JSI) of the biphotons} }
\label{Fig3}
\end{figure*}

\begin{table*}[!ht]
\centering
\caption{Comparison of the four SPDC processes. During our characterization, we fixed the pump power to 10 mW. We measure the single count (SC) and coincidence count (CC) at the degenerate temperature. We note that SC is calculated as $\rm{SC}=\sqrt{SC1\times SC2}$, where SC1 and SC2 are the single counts of two different channels. The CC is the original data without correcting the factor of 1/2 in the usage of FBS.  Here, $d(z)$ is the nonlinear coefficient of the PPKTP crystal. }
\resizebox{1.8\columnwidth}{!}{%
\begin{tabular}{c|cccc}
\hline \hline
\multicolumn{1}{c|}{Poling period} & \multicolumn{4}{c}{10 $\mu$m}                                                                                 \\ \hline
Pump wavelength                & \multicolumn{1}{c|}{404.3 nm}    & \multicolumn{1}{c|}{404.3 nm}    & \multicolumn{1}{c|}{408.8 nm}     & 315 nm    \\ \hline
Phase-matching                      & \multicolumn{1}{c|}{Type-II}  & \multicolumn{1}{c|}{Type-I}   & \multicolumn{1}{c|}{Type-0}    & Type-II \\ \hline
Polarization                        & \multicolumn{1}{c|}{H$\rightarrow$H+V}        & \multicolumn{1}{c|}{V$\rightarrow$H+H}        & \multicolumn{1}{c|}{V$\rightarrow$V+V}         & H$\rightarrow$H+V      \\ \hline
Poling order                        & \multicolumn{1}{c|}{$1^{\text{st}}$}        & \multicolumn{1}{c|}{$5^{\text{th}}$}        & \multicolumn{1}{c|}{$3^{\text{rd}}$}         & $5^{\text{th}}$       \\ \hline
Tuning range                        & \multicolumn{1}{c|}{30 $^{\circ} \mathrm{C}$ $\rightarrow$ 150 $^{\circ} \mathrm{C}$}       & \multicolumn{1}{c|}{50 $^{\circ} \mathrm{C}$ $\rightarrow$ 70 $^{\circ} \mathrm{C}$}       & \multicolumn{1}{c|}{ 50 $^{\circ} \mathrm{C}$ $\rightarrow$ 120 $^{\circ} \mathrm{C}$}        & 30 $^{\circ} \mathrm{C}$ $\rightarrow$ 150 $^{\circ} \mathrm{C}$      \\ \hline
Degenerate temperature              & \multicolumn{1}{c|}{126.8 $^{\circ} \mathrm{C}$}    & \multicolumn{1}{c|}{57.2 $^{\circ} \mathrm{C}$}     & \multicolumn{1}{c|}{58.4 $^{\circ} \mathrm{C}$}      & 62.8 $^{\circ} \mathrm{C}$    \\ \hline
SC/CC (cps)                         & \multicolumn{1}{c|}{1.3M/45k} & \multicolumn{1}{c|}{385k/15k} & \multicolumn{1}{c|}{5.4M/167k} & N/A    \\ \hline
Theoretical $d(z)$ (pm/V)                         & \multicolumn{1}{c|}{$2 d_{24} / \pi=2.32$}        & \multicolumn{1}{c|}{$2 d_{32} / 5\pi=0.55$}        & \multicolumn{1}{c|}{$2 d_{33} / 3\pi=3.59$}         & $2 d_{24} / 5\pi=0.46$       \\ \hline \hline
\end{tabular}
}
\label{Tab:2}
\end{table*}

We experimentally verified the existence of these controllable transitions of phase-matching conditions using the experimental setup shown in Fig.\,\ref{Fig2}. We have used two lasers as our pump. The first laser is a narrow-band laser (Kunteng QTechnics) with a tunable center wavelengths of 401-409 nm and a line width of 10 MHz. 
The second laser is a broadband laser \cite{Cai2022JosaB}, which has a strong-power portion at 405 nm  and a weak-power portion at around 315 nm.
During our experiment, we control the pump power using a half-wave plate (HWP) and a polarizing beam splitter (PBS). By adding another HWP after the PBS, we can also control the polarization of the pump beam. Then the pump laser is filtered by a short-pass filter and then focused by a lens (L1, with a focal length of f = 50 mm) onto a temperature controlled 10-mm-long PPKTP crystal. Our PPKTP crystal has a poling period of 10 $\mu$m, which is originally designed for a type-II phase-matched SPDC at 810 nm. The down-converted biphotons are collimated by the second lens (L2, f = 50 mm) and then filtered by a dichroic mirror and a long-pass filter. Then, these biphotons are separated by the PBS and 50:50 fiber beam splitter (FBS) which directs the photons to two avalanche photodiodes (APDs). Finally, we analyze the properties of these SPDC processes using a time-interval analyzer (Picoharp 300, PicoQuant) for single and coincidence counts. Moreover, the spectra of these biphotons are measured using a single-photon level spectrometer (SP2300, Princeton Instrument).

We report the measured single counts (SC) and coincidence counts (CC) in Tab.\,\ref{Tab:2}. We also plot the experimentally measured spectra in Fig.\,\ref{Fig3}(a-d). For the $1^\textrm{st}$-order type-II case, the pump wavelength is set at 404.3 nm, and the polarization is set to be horizontal. In this case, the signal and idler are separated by PBS and then coupled into single-mode fibers, which are connected to two APDs. At the spectral degeneracy temperature of 126.8 $^{\circ} \mathrm{C}$, SC and CC are measured to be 1.3 Mcps and 45 kcps, respectively. By adjusting the temperature of the crystal from 30 $^{\circ} \mathrm{C}$ to 150 $^{\circ} \mathrm{C}$, we measured the spectra shown in Fig.\,\ref{Fig3}(a). The center wavelength of the signal and idler photon, as a function of temperature, also known as the tuning curve, is shown in Fig.\,\ref{Fig3}(e). 

For the $5^\textrm{th}$-order type-I case, we kept the pump wavelength at 404.3 nm, but we change the polarization of the pump to vertical. In this case, the signal and idler have the same horizontal polarization (H-polarized). These generated biphotons will transmit through the PBS, and we separate them using a  50:50 FBS to measure SC and CC. In particular, the measured SC and CC are 385 kcps and 15 kcps at the spectral degeneracy temperature of 57.2 $^{\circ} \mathrm{C}$. By adjusting the temperature from 50 $^{\circ} \mathrm{C}$ to 90 $^{\circ} \mathrm{C}$, we measured the spectra shown in Fig.\,\ref{Fig3}(b). The tuning curve is shown in Fig.\,\ref{Fig3}(f), which is clearly different from Fig.\,\ref{Fig3}(e).

For $3^\textrm{rd}$-order type-0 case, the pump wavelength is set to 408.8 nm, and the polarization is set to be vertical. The generated vertically-polarized biphotons are reflected by the PBS, collected and separated by another 50:50 FBS, and finally measured by two APDs. At the spectral degeneracy temperature of 58.4 $^{\circ} \mathrm{C}$, the SC and CC are 5.4 Mcps and 167 kcps, respectively. In this case, we tune the temperature of the PPKTP crystal from 50 $^{\circ} \mathrm{C}$ to 120 $^{\circ} \mathrm{C}$, and we show the measured spectra in Fig.\,\ref{Fig3}(c). The corresponding tuning curve is shown in Fig.\,\ref{Fig3}(g).

In our final measurement, we use a broadband laser with a central wavelength of 315 nm to implement the $5^\textrm{th}$-order type-II SPDC process. We note that in this case, the pump power is unknown due to the limitations of our devices. Nevertheless, in this case, the measured spectra and tuning curve are shown in Fig.\,\ref{Fig3}(d) and (h). The measured temperature range is 30 $^{\circ} \mathrm{C}$ to 150 $^{\circ} \mathrm{C}$ and the spectral degeneracy temperature is measured to be 62.8 $^{\circ} \mathrm{C}$. 
{The observed wavelength of 315 nm slightly deviates from the theoretically calculated value of 311.43 nm, which may indicate that the accuracy of the Sellmeier equation used in the calculations needs further improvement. }

As a general observation, we note that the type-II SPDC has a much narrower FWHM than type-I and type-0 SPDC, as also indicated in Fig.\,\ref{Fig3}(i-l).

\section{Discussion}
{To gain a deeper understanding of the biphotons generated under different phase-matching condition, we calculated the joint spectral intensities (JSIs) \cite{Jin2013OE} under wavelength degenerate condition, as depicted in Fig.\,\ref{Fig3}(m-p).
Notably, the JSIs in panel (m-o) are  distributed along the anti-diagonal direction. This is attributed to the narrow spectral width of the pump laser, causing the pump envelope function to predominantly shape the JSI.
However, in panel (p), the JSI deviates from the anti-diagonal direction. This discrepancy arises from  the broader spectral width of the pump laser, causing the phase-matching function to dominate the JSI.}

{In this experiment, we only observed four SPDC processes as listed in Tab.\,\ref{Tab:2}. In the future, it is possible to observe more than nine phase-matching conditions as long as the condition in Eq.\,\ref{eq2} is satisfied.
Our biphoton source allows for both broadband (type-0 and type-I) and narrowband (type-II) biphoton emission without changing the experimental setup or PPKTP crystal, which is convenient for operations in experiment. In the future, this biphoton source can be  upgraded to a polarization-entangled photon source or a time-bin entangled photon source \cite{Zhang2021npj, Shen2023LSA}.}

{One remarkable finding of this study is the observation of three phase-matching conditions  matched within a very short pump wavelength range of 401-408 nm.
This wavelength range, particularly around 405 nm, holds significant importance for entangled photon sources, being widely utilized not only in laboratory \cite{Fedrizzi2007OE,Liu2021PR,Hong2023AQT} but also in satellite-based endeavors \cite{Yin2017Science}. 
The popularity of this wavelength range can be attributed to the maturity of blue-violet laser technology, which allows the production of high-power blue lasers at relatively affordable costs \cite{Jeong2016OE,Lohrmann2020APL,Cai2021JosaB,Yang2023OL,Wang2020PRL,Zhang2020PRL}.
The identification of three phase-matching conditions in this study holds the potential to  advance the quantum applications of entangled photon sources utilizing blue lasers and PPKTP crystals.}

\section{Conclusion}
Entangled photon pairs are crucial for quantum information processing protocols. The possibility of realizing all three types of SPDC processes in a single PPKTP crystal could be beneficial for future research. Here, we experimentally observed four SPDC processes in a single PPKTP crystal with a poling period of 10 $\mu$m. These processes include $3^\textrm{rd}$-order type-0, $1^\textrm{st}$-order type-II, $5^\textrm{th}$-order type-I, and $5^\textrm{th}$-order type-II. We have also reported on the properties of these processes, including the coincidence count and thermally dependent spectra. Our scheme can provide a versatile biphoton source for future quantum protocols, as it allows for both broadband (type-0 and type-I) and narrowband (type-II) biphoton sources without changing the experimental setup or PPKTP crystal. 

\subsection*{Appendix A: Calculation of $d_{\text{eff}}$ for KTP}
Here we derive the effective nonlinear coefficient of KTP $d_{\text{eff}}$. 
In the interaction of strong light with a nonlinear medium, the relationship between the dielectric polarization density $\vec{P}$ and the electric field $\vec{E}$ is nonlinear, and the $\vec{P}$ induced by the medium can be expanded into a power series of $\vec{E}$. We only consider the second-order case \cite{boyd2020nonlinear}:
\begin{equation}
\vec{P}(E)=\varepsilon_0 \chi^{(2)} \vec{E} \vec{E},
\end{equation}
where $\varepsilon_0$ is the permittivity of free space, $\chi^{(2)}$ is the second-order nonlinear optical susceptibility, and $\vec{E}\vec{E}$ is the tensor product of the electrical field. When the Kleiman symmetry condition is established: $d_{i j k}=\frac{1}{2} \chi_{i j k}^{(2)}$. The nonlinear polarizations of the interacting waves in a PPKTP crystal can be given in the form:
\begin{equation}
\left[\begin{array}{c}
P_x \\
P_y \\
P_z
\end{array}\right]=2 \varepsilon_0\left[\begin{array}{llllll}
d_{11} & d_{12} & d_{13} & d_{14} & d_{15} & d_{16} \\
d_{21} & d_{22} & d_{23} & d_{24} & d_{25} & d_{26} \\
d_{31} & d_{32} & d_{33} & d_{34} & d_{35} & d_{36}
\end{array}\right]\left[\begin{array}{c}
E_x^2 \\
E_y^2 \\
E_x^2 \\
2 E_y E_z \\
2 E_x E_z \\
2 E_x E_y
\end{array}\right],
\end{equation}
where $P_i$ ($i$=$x$, $y$, $z$) is the component of $\vec{P}$ along the $x-$, $y-$, and $z-$axes; $d_{ij}$ ($i=1-3$; $j=1-6$), which is simplified from $d_{i j k}$, represents the nonlinear susceptibility tensor of the KTP crystal. $E_i$ ($i=1,2,3$) is the electrical field of the fundamental wave along the $x-$, $y-$, and $z-$axes.
For type-II SPDC, 
$P_y=\varepsilon_0 d_{2 4} E_y E_z$, and $d_{\text{eff}}$=$d_{2 4}$.
For type-I SPDC,
$P_z=\varepsilon_0 d_{3 2} E_y E_y$, and $d_{\text{eff}}$=$d_{3 2}$;
For type-0 SPDC, 
$P_z=\varepsilon_0 d_{3 3} E_z E_z$, and $d_{\text{eff}}$=$d_{3 3}$.
Finally, for PPKTP, $d_{2 4}=3.64$ pm/V, $d_{3 2}=4.35$ pm/V, and $d_{3 3}=16.9$ pm/V \cite{Vanherzeele1992}.

\subsection*{Appendix B: Calculation of Fourier coefficient $G_{m}$}
Here we derive the Fourier coefficient $G_{m}$  (Eq.\,\ref{eq5}) in the main text. Considering $d(z)$ is a square-wave function:
\begin{equation}
d(z)=d_{\text{eff}} \mathrm{sign}[\cos (2 \pi z / \Lambda)],
\end{equation}
where $d_{\text{eff}}$ denotes the nonlinear coefficient of the KTP and $\Lambda$ is the poling period of PPKTP.

For arbitrary periodic signals that satisfy the Dirichlet conditions \cite{lanczos2016discourse}, they can be rewritten as Fourier series in exponential form:
\begin{equation}
f(t)=\sum_{n=-\infty}^{\infty} c_n e^{i n \omega_0 t}, n=0, \pm 1, \pm 2, \cdots,
\end{equation}
where
\begin{equation}
c_n=\frac{1}{T} \int_{-\frac{T}{2}}^{\frac{T}{2}} f(t) e^{-i n \omega_0 t} d t.
\end{equation}

Therefore, the Fourier coefficient $G_{m}$ of $d(z)$ is calculated as follow:
\begin{equation}
 \begin{aligned}
&G_m=\frac{1}{\Lambda} \int_{-\frac{\Lambda}{2}}^{\frac{\Lambda}{2}} \operatorname{sign}[\cos (2 \pi z / \Lambda)] e^{-i \frac{2 \pi m}{\Lambda} z} d z\\
&=\frac{1}{\Lambda}\int_{-\frac{\Lambda}{2}}^{-\frac{\Lambda}{4}}(-1) e^{-i \frac{2 \pi m}{\Lambda} z} d z+\frac{1}{\Lambda}\int_{-\frac{\Lambda}{4}}^{\frac{\Lambda}{4}}(1) e^{-i \frac{2 \pi m}{\Lambda} z} d z\\  &\quad + \frac{1}{\Lambda}\int_{\frac{\Lambda}{4}}^{\frac{\Lambda}{2}}(-1) e^{-i \frac{2 \pi m}{\Lambda} z} d z\\
&=\frac{1}{i 2 \pi m}\left[4 i \sin \left(\frac{\pi m}{2}\right)-2 i \sin (\pi m)\right] \\
&=\frac{2}{m\pi}{\rm{sin}}(\frac{m\pi}{2})\\
&={\rm{sinc}}(\frac{m\pi}{2}).
 \end{aligned}
\end{equation}
Now, we obtained the value of $G_{m}$ in Eq.\,\ref{eq5} of the main text.

Additionally, $d(z)$ can also be expressed in the form of square wave:\\
For $|z|<\frac{\Lambda}{2}$,
\begin{equation}
d(z)=d_{\text {eff}} \cdot\left[2 \operatorname{rect}\left(\frac{2 z}{\Lambda}\right)-1\right];
\end{equation}
and for $\left|\frac{\Lambda}{2}\right|<|z|<\left|\frac{L}{2}\right|$ (L is the length of crystal),
\begin{equation}
d(z)=d(z+m \Lambda) \quad m=0, \pm 1, \pm 2 \cdots.
\end{equation}
We note that Eq.\,\ref{eq5} can be intuitively understood as the Fourier transform of a \textit{square-wave} function to a \textit{sinc} function.

\section*{Acknowledgments}
We thank Prof. Zhi-Yuan Zhou for helpful discussion.
This work was supported by the National Natural Science Foundations of China (Grant Numbers 92365106, 12074299, and 11704290) and the Natural Science Foundation of Hubei Province (2022CFA039).

\end{document}